\documentclass{nature}
\usepackage{amsmath}
\usepackage{graphicx}
\usepackage{float}
\usepackage{ragged2e}
\usepackage{url}

\graphicspath{{ke2021_}}

\title{Earthquake Breakdown Energy Scaling Despite Constant Fracture Energy}
\author{Chun-Yu Ke$^1$, Gregory C. McLaskey$^1$, David S. Kammer$^2$}

\makeatletter
\let\saved@includegraphics\includegraphics
\AtBeginDocument{\let\includegraphics\saved@includegraphics}
\renewenvironment*{figure}{\@float{figure}}{\end@float}
\makeatother

\begin{document}

\maketitle

\begin{affiliations}
\item School of Civil and Environmental Engineering, Cornell University, Ithaca, New York, USA
\item Institute for Building Materials, ETH, Z{\"u}rich, Switzerland
\end{affiliations}

\begin{abstract}
In the quest to determine fault weakening processes that govern earthquake mechanics, it is common to infer the earthquake breakdown energy from seismological measurements. Breakdown energy is observed to scale with slip, which is often attributed to enhanced fault weakening with continued slip or at high slip rates, possibly caused by flash heating and thermal pressurization. 
However, breakdown energy varies by more than six orders of magnitude, which is physically irreconcilable with prevailing material properties. 
We present a dynamic model that demonstrates that breakdown energy scaling can occur despite constant fracture energy and does not require thermal pressurization or other enhanced weakening. Instead, earthquake breakdown energy scaling occurs simply due to scale-invariant stress drop overshoot, which is affected more directly by the overall rupture mode -- crack-like or pulse-like -- rather than from a specific slip-weakening relationship.
Our findings suggest that breakdown energy may be used to discern crack-like earthquakes from self-healing pulses with negative breakdown energy.
\end{abstract} %
\section{Introduction} %

In an earthquake, strain energy stored elastically in the Earth’s crust is quickly transformed into radiated seismic waves, new fractures and wear products, and heat. This process is controlled, to a large extent, by the fault constitutive law, which describes how a fault’s strength evolves with slip or slip rate and how much energy is dissipated or available for continued rupture. Thus, the fault constitutive law and associated fault properties are key to better understand and possibly predict earthquake mechanics. However, these properties are difficult to measure in the Earth. A number of studies tried to constrain fault constitutive behavior using seismological observations of earthquakes, and, in particular, the way that earthquake parameters scale from small to large\cite{Abercrombie2005,Tinti2005,Rice2006,Viesca2015}. One significant seismological observation, known as the \textit{breakdown energy}, is thought to be related to the slip weakening process, and is often assumed to be a proxy for fracture energy. However, breakdown energy scales with slip. Hence, larger earthquakes with more total slip appear to dissipate more energy (per unit rupture area) than small earthquakes.

If breakdown energy is equivalent to fracture energy, then from a mechanical perspective, its scaling violates common physical assumptions. Fracture energy is generally assumed to be a material or interfacial property, that only depends mildly on the rupture and its propagation, and is bounded by a finite stress drop. Hence, it does not scale over many orders of magnitude as seismologically observed. Various theories have been developed over the years to explain this discrepancy. For instance, it was suggested that frictional weakening distance could increase with increasing earthquake size\cite{Abercrombie2005}. Other studies suggested secondary weakening mechanisms such as thermal pressurization\cite{Viesca2015,Perry2020,Lambert2020} that further weaken the fault as the earthquake rupture grows larger and slip distances increase. High-velocity friction experiments that show continued weakening with cumulative slip were also offered as support. However, the fracture energy measured from those experiments appears to plateau at around $2~\textrm{m}$ of slip\cite{Nielsen2016a} while seismologically estimated breakdown energy of natural earthquakes continues beyond this limit and scales across all sizes of earthquakes\cite{Abercrombie2005}. Hence, a generally accepted theory explaining the scaling of the earthquake breakdown energy remains missing. 

A running earthquake arrests either because the rupture front enters a region of the fault that is stronger or has a more stable rheology compared to the nucleation region, \textit{i.e.}, a barrier (\textit{e.g.}, ref.\cite{Lapusta2000,Barbot2019,Perry2020,Lambert2020}), or the front enters a region with low initial stress and hence subcritical driving force\cite{Ke2021}. 
While the first scenario implies a larger fault fracture energy, the second scenario does not. 

Here we will show that the observed breakdown energy scaling does not require any complex or scale dependent fault constitutive law. We will demonstrate a simple but plausible scenario where constant fault friction but non-uniform initial stress results in the same scaling of breakdown energy. Inspired by recent observations of laboratory earthquake arrest\cite{Ke2021}, our model employs simple linear slip-weakening fault friction with \emph{constant} fracture energy, and earthquakes spontaneously arrest by propagating into unfavorable initial stress conditions. 
We solve the three-dimensional model with fully dynamic numerical simulations, and determine the breakdown energy and other seismic source parameters. The result is unambiguous: breakdown energy does not correspond to the locally imposed constant fracture energy; it results from a scale-invariant stress overshoot.

\section{Results}

\begin{figure}
\makebox[\textwidth][c]{
\includegraphics{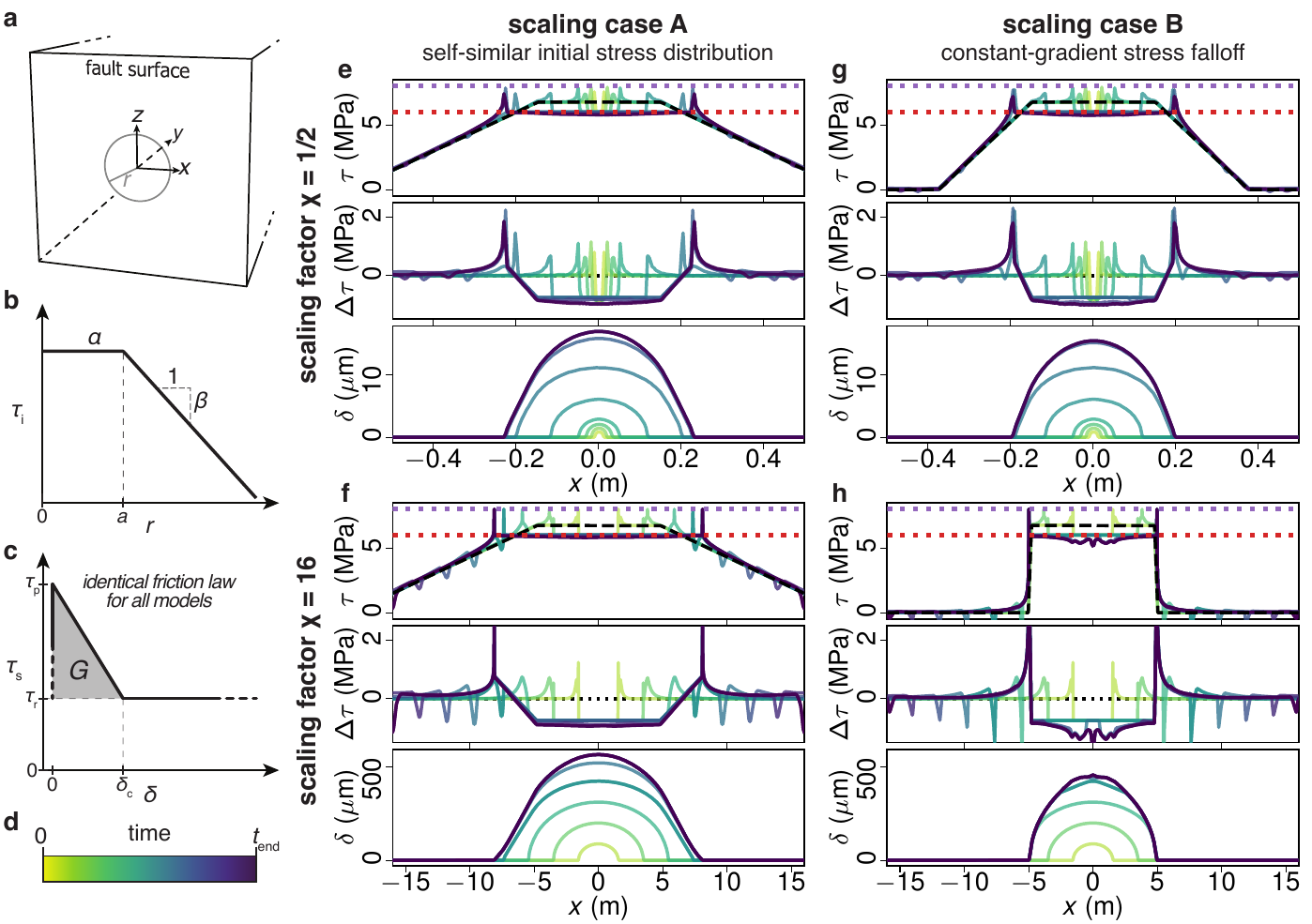}
}
\caption{Nucleation and arrest of simulated earthquake at fault with non-uniform initial stress. 
(a) Schematics of the numerical model, where the fault surface is defined on the $x$\textendash$z$ plane ($y=0$) and embedded in a homogeneous elastic full space. 
(b) Parametric initial stress distribution
where $\alpha$ is the initial stress amplitude at the plateau with radius $a$ and $\beta$ is the gradient of initial stress decrease outside the plateau. 
(c) The slip-weakening friction law where $ \tau_\mathrm{{p}}$, $ \tau_\mathrm{{r}}$, and $\delta_\mathrm{c}$ are identical for all models.
(d) The color map for curves in (e{\textendash}h) that indicates the simulation time, where $ t_\mathrm{end}$ is time of arrest.
(e{\textendash}h) are snapshots of stress $\tau$, stress drop $\Delta \tau$ and slip $\delta$ at $y=z=0$ with $\chi=2^{-1}$ and $\chi=2^4$ in scaling case A and B, respectively. 
Purple and red dotted lines indicate $ \tau_\mathrm{{p}}$ and $ \tau_\mathrm{{r}}$ from the friction law (see c), respectively. 
}\label{fig:simulation} %
\end{figure}

We consider a two-dimensional planar fault ($y=0$) embedded in a three-dimensional isotropic elastic medium with shear modulus $\mu = 12~\mathrm{GPa}$ and shear wave speed $c_\mathrm{s} = 2,126~\mathrm{m/s}$ (Fig.~\ref{fig:simulation}a). 
The fault is governed by a linear slip-weakening friction law (Fig.~\ref{fig:simulation}c) with peak strength $ \tau_\mathrm{{p}} = 8~\mathrm{MPa}$, residual strength $ \tau_\mathrm{{r}}=6~\mathrm{MPa}$ and critical slip distance $\delta_\mathrm{c} = 10^{-6}~\mathrm{m}$. Hence, the fault is characterized by a well-defined and constant fracture energy $G = 1~\mathrm{Jm^{-2}}$, consistent with previous experimental estimations\cite{Kammer2019}.
We explore two different scenarios for earthquake scaling by imposing an initial stress distribution that depends on scaling factor $\chi$. In both scaling cases, the initial stress has uniform amplitude $\alpha = 6.75~\mathrm{MPa}$ within a circular region of radius $a=0.3\chi~\mathrm{m}$, centered on the origin, \textit{i.e.}, $(x, y, z) = (0,0,0)$. 
Earthquake ruptures are nucleated at the origin by a slowly increasing area of reduced fault strength (see Supplementary Text S1). While the nucleation process is artificial, it is sufficiently small and slow to not affect the unstable propagation of the earthquake that occurs spontaneously. After nucleation, the earthquake rupture velocity quickly accelerates to the Rayleigh wave speed where it remains constant\cite{Svetlizky2017a,Chounet2018}. The earthquake is nearly axisymmetric but grows slightly faster in the direction of the applied shear load (\textit{i.e.}, $x${\textendash}direction).

The two scaling cases produce identical initial rupture growth, but they differ in the way they arrest, and this provides insight into the role that arrest plays in calculated source parameters.
In scaling case A, initial stress decreases outside the circular region (Fig.~\ref{fig:simulation}b) at spatial rate $\beta=7.5/\chi~\mathrm{MPa/m}$ (\textit{cf.} Fig.~\ref{fig:simulation}e and Fig.~\ref{fig:simulation}f and note scaling of $x$\textendash axis). In scaling case B, we impose a scale-invariant stress falloff $\beta=30~\mathrm{MPa/m}$ (\textit{cf.} Fig.~\ref{fig:simulation}g and Fig.~\ref{fig:simulation}h). The $\chi=2^{-2}$ model is identical in both scaling cases and $\chi=1$ corresponds to the scale of large laboratory earthquakes\cite{Ke2018,Ke2021}. We highlight that the resolution of the numerical models, the nucleation procedure, and fault friction properties, including fracture energy $G$  (Fig.~\ref{fig:simulation}c), are kept constant across all scales.

\subsection{Scaling of earthquake source properties} %

\begin{figure}
\centering
\includegraphics{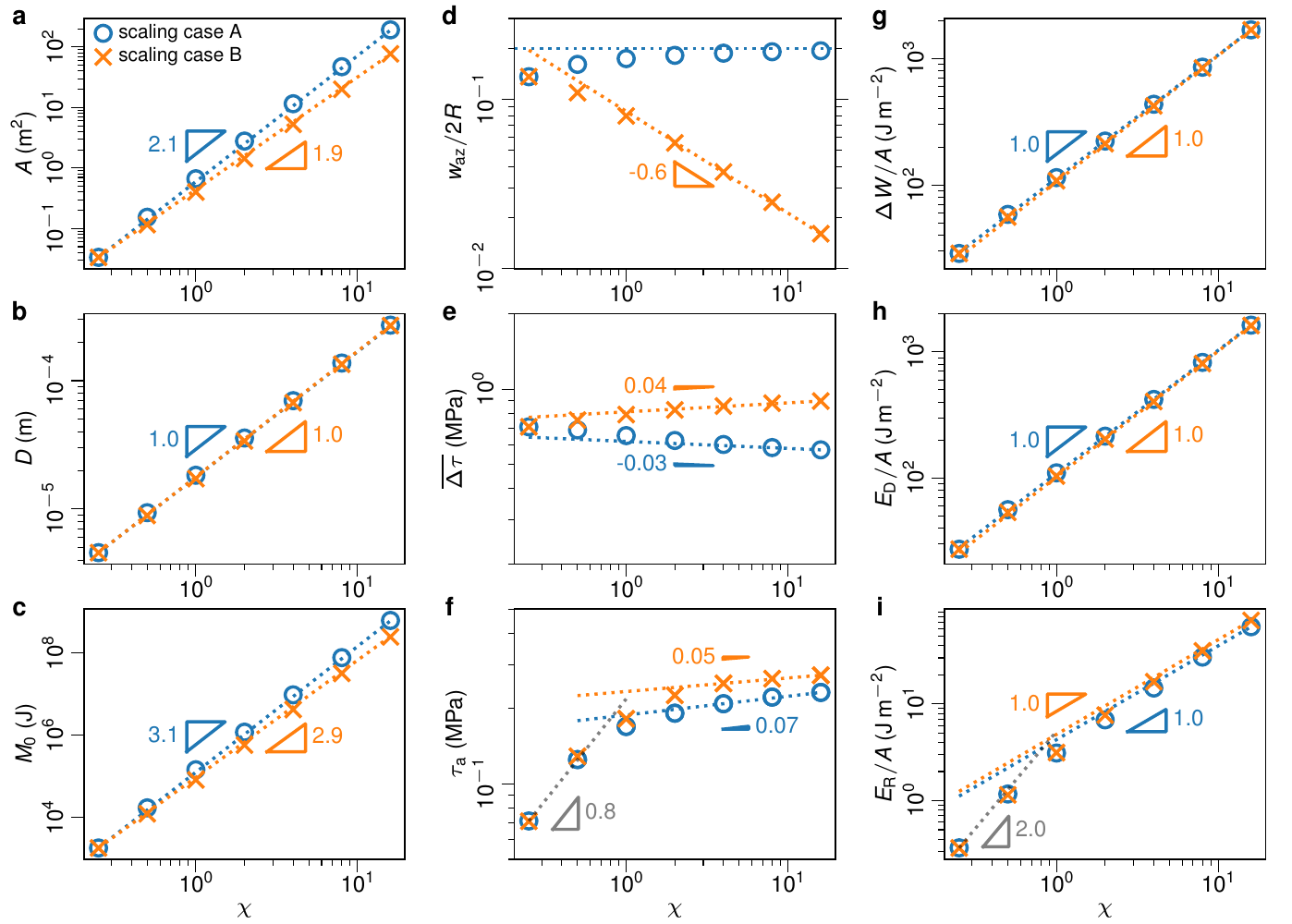}
\caption{Scaling of earthquake source parameters in both scaling cases. 
Triangles and the annotated numbers indicate the power of trend lines, \textit{i.e.}, $p$ in $V \propto \chi^p$ for different parameters $V$. 
Scaling is shown for (a)~earthquake rupture area, (b)~average slip, (c)~seismic moment, (d)~earthquake arrest zone width scaled by rupture dimension, (e)~averaged stress drop, (f)~apparent stress, (g)~averaged total released strain energy, (h)~averaged dissipated energy, and (i)~averaged radiated energy computed from integral evaluation.
}\label{fig:scaling} %
\end{figure}

The scaling of calculated earthquake source properties is shown in Fig.~\ref{fig:scaling}, and definitions for these parameters in the context of our models are describe below. Earthquake rupture area $A$ is the area where fault slip $\delta > 0$, \textit{i.e.}, $A=\int_\Sigma \mathop{}\!\mathrm{d} S$ for $\Sigma=\{x, z \in U | \delta(x, z) > 0\}$ and $U$ is the entire simulation fault surface domain with $y=0$. The spatially averaged slip distance $ D = \int_\Sigma \delta_\mathrm{f}(x, z) \mathop{}\!\mathrm{d} S / A$
where $\delta_\mathrm{f}(x, z) = \delta(x, z, t= t_\mathrm{end})$ is the final slip distance after earthquake arrest (Fig.~\ref{fig:simulation}e{\textendash}h).
Moment release rate 
$\dot{M}(t) = \mu\int_U \dot{\delta}(x, z, t)\mathop{}\!\mathrm{d} S$~, where $\dot{\delta} = \mathrm{d}\delta/\mathrm{d}t$ is the on-fault slip rate. The seismic moment is hence given by $M_0 = \int_0^{ t_\mathrm{end}} \dot{M}(t)\mathop{}\!\mathrm{d} t$.
Alternatively, the seismic moment could be computed as $M_0 = \mu A  D$, which yields equivalent results.

The averaged stress drop $\overline{\Delta\tau}^\mathrm{{}}$ is one of the most important inferred earthquake source properties. 
A common seismological approach to determine $\overline{\Delta\tau}^\mathrm{{}}$ is to assume a uniform stress drop\cite{Eshelby1957} and estimate it from seismic source parameters using $\overline{\Delta\tau}^\mathrm{{}} = 7M_0/16R^3$.
Despite the simplification of assumed uniform stress drop, which is clearly not satisfied in our model (see $\Delta\tau$ in Fig.~\ref{fig:simulation}e{\textendash}h), the above formulation provides an accurate estimation of $\overline{\Delta\tau}^\mathrm{{}}$, and more sophisticated approaches\cite{Noda2012,Noda2013} do not lead to significantly different results (see Supplementary Text S2).

We observe that $\overline{\Delta\tau}^\mathrm{{}}$ is nearly scale-invariant in our model (Fig.~\ref{fig:scaling}e), where scaling case A and B bound closely a scale-invariant behavior, consistent with seismologically inferred measurements\cite{Ide2001,Prieto2004,Allmann2009,Yoshimitsu2014}. 
With the self-similar initial stress distribution of scaling case A, the arrest zone $ w_\mathrm{az}$\cite{Ke2021} is a nearly constant percentage of the average rupture radius $R$ (Fig.~\ref{fig:scaling}d). (We define $ w_\mathrm{az} = R - a$, where $R=\sqrt{A/\pi}$ and $A$ is the rupture area.) For scaling case A, initial stress  decreases more gradually (smaller $\beta$) for larger earthquakes than for smaller ones, and $A$ scales with $\chi^{2.1}$ rather than the expected $\chi^{2.0}$ because large earthquakes rupture proportionally further into unfavorably stressed regions. For this reason $M_0$ scales with $\chi^{3.1}$ instead of $\chi^{3.0}$. However, the average slip $ D$ scales as $\chi^{1.0}$, therefore stress drop ($\sim D/\sqrt{A}$) decreases slightly (Fig.~\ref{fig:scaling}e). 
In scaling case B, $ w_\mathrm{az}/2R$ is smaller for larger earthquakes (Fig.~\ref{fig:scaling}d) and $A$ scales as $\chi^{1.9}$ (rather than $\chi^{2.0}$) because larger and larger ruptures meet relatively steeper and steeper stress gradients and therefore rupture proportionally shorter and shorter distances into unfavorable stress. Similarly, $M_0$ scales with $\chi^{2.9}$ and stress drop increases slightly with increasing earthquake size. 

We also analyze the energy budget for our model earthquakes. Using the energy-considered averaged initial stress $\overline{{{\tau}}}_\mathrm{{i}}$ and final stress $\overline{{{\tau}}}_\mathrm{{f}}$\cite{Noda2012,Noda2013}, the averaged total released strain energy is computed $\Delta W / A= \frac{1}{2} \left(\overline{{{\tau}}}_\mathrm{{i}} + \overline{{{\tau}}}_\mathrm{{f}}\right) D$, which is a simplified formulation of the method proposed by ~ref.\cite{Kostr1976} (see Fig.~\ref{fig:scaling}g and Methods).
We also compute the dissipated energy $ E_\mathrm{{D}} =  E_\mathrm{{H}} +  E_\mathrm{{G}}$, which combines heat $ E_\mathrm{{H}}$ and total breakdown energy $ E_\mathrm{{G}}$, by
$ E_\mathrm{{D}} = \int_0^{ t_\mathrm{end}}\int_U \tau(x, z, t)\dot{\delta}(x, z, t) \mathop{}\!\mathrm{d} S \mathop{}\!\mathrm{d} t$ (Fig.~\ref{fig:scaling}h). Finally, we estimate the radiated energy $ E_\mathrm{{R}}$ (see Methods and Supplementary Text S3),  
and we observe $ E_\mathrm{{R}}/A \propto  D^1$, consistent with a nearly constant apparent stress $ \tau_\mathrm{{a}} = \mu E_\mathrm{{R}}/M_0 = 100{\text\textendash}200~\mathrm{kPa}$ (Fig.~\ref{fig:scaling}f,i; ~ref.\cite{Ide2001,Convers2011,Baltay2014,Denolle2016}).

From our scaling analysis, we conclude that our model reproduces the scaling behavior of all important seismic source properties. The expected scaling properties for self-similar rupture are either reproduced by the models or are bounded by scaling cases A and B (Fig.~\ref{fig:scaling}a{\textendash}d). We specifically note that $A \propto  D^2$, $M_0 \propto  D^3$, and $\overline{\Delta\tau}^\mathrm{{}} \propto  D^{0}$
, consistent with standard earthquake scaling\cite{Walter2006}. 

\subsection{Scaling of breakdown energy} %

\begin{figure}
\makebox[\textwidth][c]{
\includegraphics{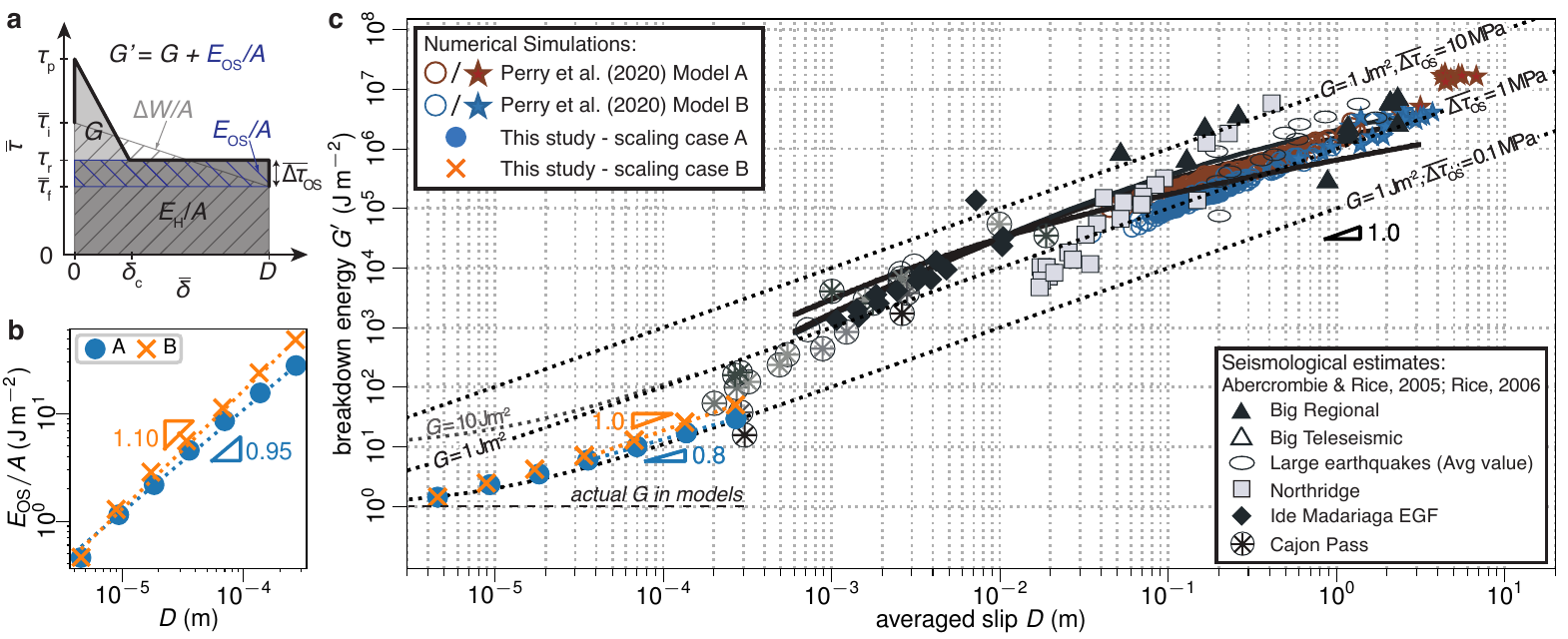}
}\caption{Scaling relations of seismologically estimated breakdown energy $ G^\prime$. 
(a)~Schematics of energy partition on the $\overline{{\tau}}$\textendash$\overline{{\delta}}$ space, where the dark blue hashed area indicates the overestimation of fracture energy by $ G^\prime$ due to stress overshoot, \textit{i.e.}, $\overline{{{\tau}}}_\mathrm{{f}}< \tau_\mathrm{{r}}$. 
(b)~The overshoot energy $ E_\mathrm{{OS}}/A$ (Eqn.~\ref{eqn:EO}) is equivalent to $ G^\prime-G$, which scales with $ D^1$.
(c)~Comparison of $ G^\prime$ from our models to data by ref.\cite{{Abercrombie2005},{Rice2006},{Perry2020}}. Three black dotted curves indicate $ G^\prime$ computed by Eqn.~\ref{eqn:Gp2} with $G=1~\mathrm{Jm^{-2}}$. The gray dotted curve diverged from the $\overline{\Delta\tau}_\mathrm{OS}=1~\mathrm{MPa}$ curve at the lower end of $ D$ demonstrates that $G$ is the lower limit of $ G^\prime$ for $\overline{\Delta\tau}_\mathrm{OS} \geq 0$.
}\label{fig:scalingG} %
\end{figure}

The dissipated energy $ E_\mathrm{{D}}$ is composed of the total breakdown energy $ E_\mathrm{{G}}$ and heat $ E_\mathrm{{H}}$ (Fig.~\ref{fig:scalingG}a). 
The averaged breakdown energy $\overline{G} \approx  E_\mathrm{{G}}/A$ is often interpreted as a proxy for the fracture energy $G$, which controls the dynamics of the fault rupture\cite{Svetlizky2017a}, and hence is of great importance for seismology.
~ref.\cite{Abercrombie2005} proposed a seismological approach to estimate $\overline{G}$ by
\begin{equation}\label{eqn:Gp}
     G^\prime = \frac{ D}{2} \left(\overline{\Delta\tau}^\mathrm{{}}-\frac{2\mu E_\mathrm{{R}}}{M_0}\right)\ ,
\end{equation}
which has been widely used (\textit{e.g.}, ref.\cite{Antolik2004,Beeler2006,Prieto2013,Ye2016,Denolle2016}).
We compute $ G^\prime$ for our simulations following Eqn.~\ref{eqn:Gp} and observe that our scaling cases A and B follow $ G^\prime\propto D^{0.8}$ and $ G^\prime\propto D^{1.0}$, respectively (Fig.~\ref{fig:scalingG}c). 
~ref.\cite{Abercrombie2005} found $ G^\prime\propto D^{1.28}$ . Thus, our models show a similar scaling of $ G^\prime$, even though the actual fault property of fracture energy $G$ is constant across all scales in our simulations (Fig.~\ref{fig:scalingG}). 

Using energy calculations proposed by ref.\cite{{Noda2012},{Noda2013}}, we rewrite $ G^\prime$ as 
\begin{equation}\label{eqn:Gp2}
 G^\prime=G + \overline{\Delta\tau}_\mathrm{OS}  D~,
\end{equation}
where $\overline{\Delta\tau}_\mathrm{OS}=\left( \tau_\mathrm{{r}} - \overline{{{\tau}}}_\mathrm{{f}}\right)$ is the energy-considered averaged stress overshoot (see Fig.~\ref{fig:scalingG}a and Methods). Hence, the breakdown energy consists of the sum of the fracture energy and the overshoot energy, which we define as
\begin{equation}\label{eqn:EO}
 E_\mathrm{{OS}}/A\equiv\overline{\Delta\tau}_\mathrm{OS} D~.
\end{equation}
In our simulations, the averaged stress overshoot $\overline{\Delta\tau}_\mathrm{OS} = 100{\text\textendash}200~\mathrm{kPa}$ is positive and nearly scale-invariant, and the overshoot energy scales as $ E_\mathrm{{OS}}/A\propto D^1$ (Fig.~\ref{fig:scalingG}b). 
Consequently, $ G^\prime$ of large earthquakes with small $G$ and scale invariant overshoot is dominated by the overshoot, hence: $ G^\prime\propto D^{1}$. 
Stress overshoot occurs in crack-like ruptures when a fast-propagating rupture front arrests but parts of the fault continue to slip\cite{Madariaga1976}, \textit{i.e.}, dynamic overshoot. With Eqn.~\ref{eqn:Gp2}, we can infer the stress overshoot of natural faults (see Fig.~\ref{fig:scalingG}c), and find $\overline{\Delta\tau}_\mathrm{OS} \approx 1~\mathrm{MPa}$. 
Further, $G$ serves as a lower bound of $ G^\prime$, as shown by the dotted curves at lower $ D$ in Fig.~\ref{fig:scalingG}c. This implies the fault fracture energy is bound by $G \leq 10~\mathrm{Jm}^{-2}$. 

If our interpretation is correct, and $G$ is indeed small ($G \leq 10~\mathrm{Jm}^{-2}$), how can this be reconciled with estimates of $G ~= 10^6~\mathrm{Jm}^{-2}$ derived from finite-fault kinematic inversions (\textit{e.g.}, ref.\cite{Tinti2005}), or with laboratory data that indicates $1~\mathrm{m}$ weakening distances\cite{Nielsen2016}? 
As for the kinematic inversions, we find that the minimum resolvable characteristic weakening length $\hat \delta_\mathrm{c}$ could be limited by bandwidth\cite{Guatteri2000}. For example, assuming a typical $0.5~\mathrm{s}$ smoothing operator (\textit{e.g.}, ref.\cite{Tinti2005}), minimum resolvable  $\delta_\mathrm{c}$ is of order $500~\mathrm{mm}$. Assuming $ \tau_\mathrm{{p}}- \tau_\mathrm{{r}}=10~\mathrm{MPa}$, this places minimum resolvable $G = 2.5~\mathrm{MJ/m^2}$. (See Supplementary Text~S4, Fig.~S4.) Considering the laboratory results, most of the observed weakening occurs while fault slip accelerates\cite{Nielsen2016}, and $1~\mathrm{m}$ weakening distances resulted from sluggish slip acceleration ($6.5~\mathrm{m/s^2}$) compared to measurements from dynamic rupture fronts ($>20,000~\mathrm{m/s^2}$)\cite{McLaskey2015}. Experiments that imposed more abrupt loading ($30~\mathrm{m/s^2}$) exhibited far smaller weakening distances ($0.02~\mathrm{m}$)\cite{Liao2014}. Thus, large inferred weakening distances and therefore large $G$ may be an artifact of laboratory loading procedures that are slow compared to the fault acceleration imposed by a dynamic rupture front during an earthquake.

Finally, our interpretation helps reconcile observations of $ G^\prime \approx 0$ and negative $ G^\prime$, found for numerous earthquakes\cite{Abercrombie2005,Viesca2015}, yet rarely discussed. Such cases would result from negligible overshoot or undershoot \textit{i.e.}, $\overline{{{\tau}}}_\mathrm{{f}} >  \tau_\mathrm{{r}}$, a condition typically associated with pulse-like earthquake ruptures\cite{Lambert2021}, where the fault is elastically reloaded after a short period of slip. Indeed, we find that a simple model with scale-invariant random overshoot that ranges between $-1$ and $2~\mathrm{MPa}$ and negligibly small $G$ produces $ G^\prime\propto D^{1.0}$, fits the ~ref.\cite{Abercrombie2005} data reasonably well (see Fig.~S12 and dotted curves in Fig.~\ref{fig:scalingG}c), and produces a catalog where one third of all events exhibit $ G^\prime\leq 0$, similar to the large earthquakes studied by ~ref.\cite{Viesca2015}. All in all, our simulations and proposed interpretation of $ G^\prime$ reconciles the co-existence of breakdown energy scaling and negative $ G^\prime$.

\subsection{Rupture growth and arrest from source time functions} %

\begin{figure}
\centering
\includegraphics{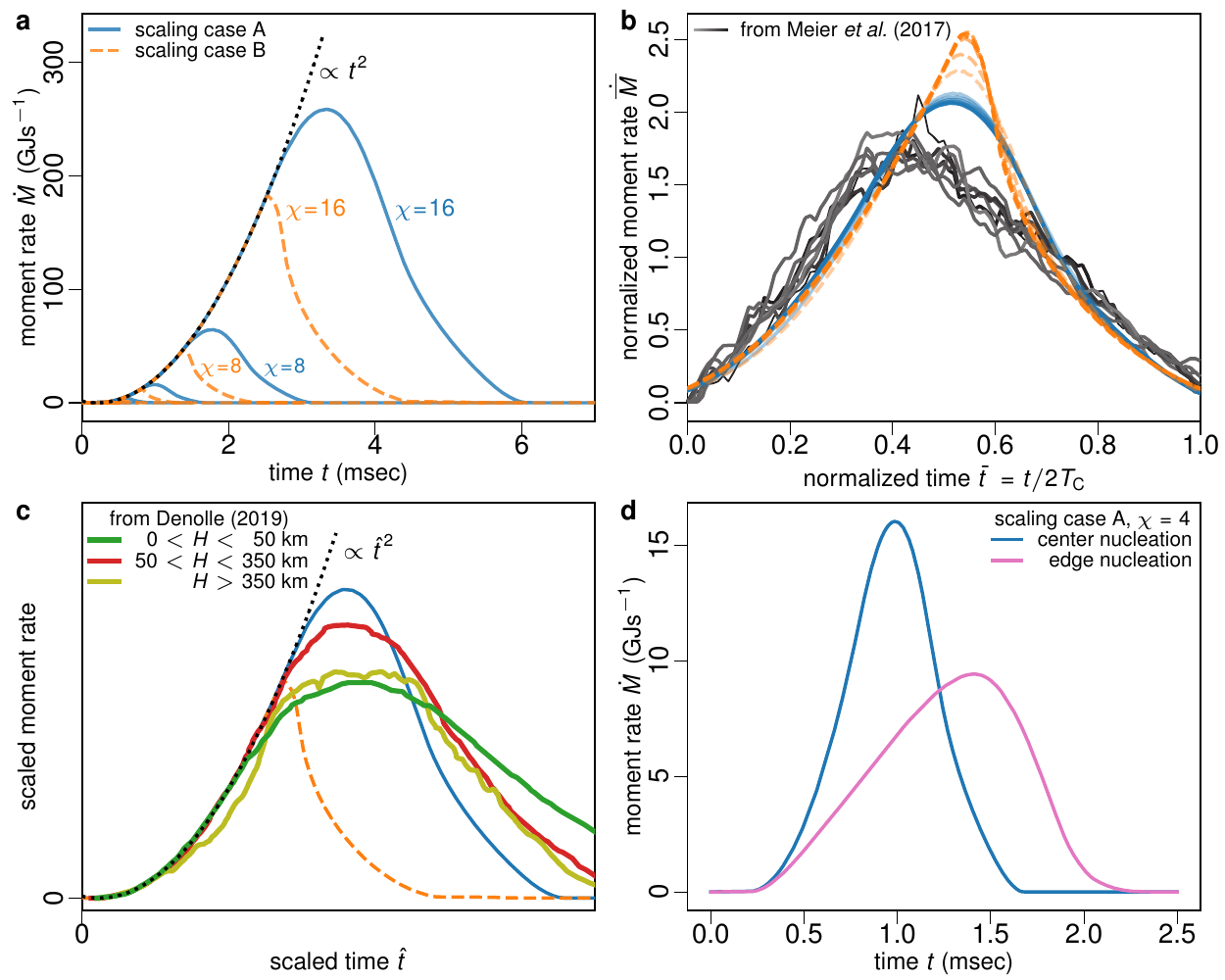}
\caption{Source time function $\dot{M}(t)$ of numerical simulations and natural earthquakes. 
(a) All models follow an identical $\dot{M}(t)\propto t^2$ growth independent of $\chi$ until reaching unfavourable stress conditions. Dotted curve is quadratic fit for growth phase of moment rate history.
(b) Normalized source time functions, where $T_\mathrm{C}=\int_0^\infty t\dot{M}\mathop{}\!\mathrm{d} t / \int_0^\infty \dot{M}\mathop{}\!\mathrm{d} t $ and the area under $\dot{\overline{M}}(\overline{t})$ is 1. Gray curves are natural earthquakes  with $M_\mathrm{w} \geq 7$ presented in ~ref.\cite{Meier2017}.
(c) Scaled source time functions with matching growth phase. Green and red curves are natural earthquakes adapted from ~ref.\cite{Denolle2019}. 
(d) Source time function of a rupture nucleated at the center of the stress plateau, and at the edge of the stress plateau.
}\label{fig:Mdot} %
\end{figure}

We study the spontaneous growth and arrest of rupture in our dynamic model by comparing the earthquake source time function or moment rate function $\dot{M}(t)$ with natural observations. $\dot{M}(t)$
consists of three phases: (1) a self-similar growth phase (2) a divergence from self-similar growth near peak moment rate and (3) the post-peak decay. In the growth phase, we observe $\dot{M}(t)\propto t^2$, at all scales (see Fig.~\ref{fig:Mdot}a), as expected for roughly circular earthquake ruptures propagating without bound\cite{Sato1973,Uchide2010}. Self-similar moment rate growth has also been observed for large earthquakes -- in some cases $\dot{M}(t)\propto t^2$ (Fig.~\ref{fig:Mdot}c; ~ref.\cite{Denolle2019}) but in others with $\dot{M}(t)\propto t^1$ (Fig.~\ref{fig:Mdot}b; ~ref.\cite{Meier2017}). Near peak $\dot{M}(t)$, earthquake rupture begins to arrest and the boundaries to rupture growth are increasingly felt. Scaling case B produces earthquakes that arrest far too quickly compared to observations\cite{Meier2017,Denolle2019}. Scaling case A's gradual arrest is a reasonable fit to observations, however, all of our models exhibit rapid post-peak decay of $\dot{M}(t)$ and negative skew, while natural earthquakes show a more gradual post-peak behavior and slightly positive skew\cite{Meier2017,Denolle2019}. This suggests that our simulated earthquakes do not arrest slowly enough or that slip ceases too quickly after the rupture initially begins to arrest. 

To better understand the above discrepancies, we compared the previously described simulations, where nucleation occurs in the center of the circular region of favorable initial stress, to a case where the earthquake nucleates close to the edge of the favorably stressed region, \textit{i.e.}, $(x, y, z)=(0.7a, 0, 0.7a)$, denoted edge nucleation (see Fig.~S11). The latter case quickly reaches unfavorable initial stress on NE side and then ruptures unilaterally to the SW. The difference in the source time function is striking (Fig.~\ref{fig:Mdot}d). The edge nucleation quickly diverges from the $\dot{M}(t)\propto t^2$ self-similar growth curve and instead $\dot{M}(t)$ grows nearly linearly. The growth phase is about 50 percent longer than in the symmetric case, though the decay is very similar. 

While the edge nucleation model does not reconcile the abbreviated post-peak behavior (and actually makes the skewness worse) it offers a satisfying explanation for linear growth rate. The earliest part of the growth phase, when unbounded growth is expected, is difficult to resolve from kinematic models\cite{Meier2017}. Proper resolution of early growth would require fault plane discretization size to depend on the distance from the hypocenter. When normalized source time functions are plotted on a linear scale, as shown in Fig.~\ref{fig:Mdot}b (or ~ref.\cite{Meier2017} Fig.~2B-D), their form is dominated by the final increase in moment rate just before rupture arrest exceeds rupture growth. In this stage, rupture growth will, on average, be bounded on at least one side, and will produce the nearly linear increase in moment rate demonstrated by our edge nucleation model.

\section{Discussion}

Previous work assumed that breakdown energy $ G^\prime$ was a proxy for fracture energy $G$, and was therefore dominated by the way strength evolved with slip. Our modelling offers an alternative interpretation. It demonstrates that scaling of $ G^\prime$ can arise naturally from the dynamics of rupture growth and arrest, and it can occur even with constant $G$ and does not require a scale-dependent constitutive behavior such as enhanced weakening at large fault slip.  
As noted by ~ref.\cite{Abercrombie2005}, $ G^\prime$ depends strongly on overshoot and undershoot, which are affected by the rupture type (crack versus pulse) and the manner in which an earthquake arrests. Our observation of $ G^\prime$ scaling is directly linked to the fact that our model exhibits scale-independent stress overshoot. 
This also suggests that negative $ G^\prime$ is an indication of undershoot, which is thought to occur for self-healing slip pulses, while the majority of earthquakes overshoot\cite{Beeler2003}, a property of crack-like ruptures. However, our models’ overshoot-induced scaling ($ G^\prime\propto D^{0.8}$ and $ G^\prime\propto D^{1.0}$) is somewhat weaker than the $ G^\prime\propto D^{1.28}$ observed by ~ref.\cite{Abercrombie2005}. Our model produces a self-similar source time function and offers an explanation for the seismologically observed linear growth phase\cite{Meier2017}, but decays too quickly compared to natural observations.  

\clearpage
\section*{Methods}
\label{sec:method}
\setcounter{equation}{0} \renewcommand{\theequation}{M\arabic{equation}} 

\subsection*{Parametrization of Initial Stress Distribution}

The on-fault initial stress distribution $ \tau_\mathrm{{i}}(x, y=0, z)$ is applied in the x-direction and is parameterized as 
\begin{equation}\label{eqn:tau_i}
 \tau_\mathrm{{i}}(r) = \alpha - \beta (r - a) H(r - a) ~,
\end{equation}
where $r=\sqrt{x^2+z^2}$ is the distance to the point of nucleation, $H$ is the Heaviside step function, $a$ is the radius of the stress plateau with amplitude $\alpha$, and $\beta$ is the spatial-rate of stress decay outside the stress plateau (Fig.~\ref{fig:simulation}b). 

\subsection*{Numerical Method}
Numerical simulations are conducted with our implementation of the spectral boundary integral (SBI) method\cite{uguca} to solve for a fully dynamic interface debonding problem. The method is based on previous methodological studies\cite{Geubelle1995,Geubelle1997,Breitenfeld1998} and the implementation has been verified through comparison with SCEC-USGS dynamic rupture code verification examples\cite{Harris2018}.
The SBI method has inherent periodic boundary conditions at the boundaries of the simulated fault domain, \textit{i.e.}, $(x, y, z) = (\pm L/2, 0, \pm L/2)$ in domain $\{x \in [-L/2, L/2); y = 0; z \in [-L/2, L/2)\}$. 
We verified that the simulated domain is large enough that the rupture and the reflected waves do not affect the ruptured region within the simulation duration. 
Ruptures were nucleated by a seed crack at the center of the fault, in which the peak strength is manually decreased to $ \tau_\mathrm{{r}}$ and extended at 10\% of the Rayleigh wave speed (see Supplementary Text S1). 

\subsection*{Seismic Source Parameters}
$M_0$ calculated through integration over $\dot{M}(t) = \mu\int_U \dot{\delta}(x, z, t)\mathop{}\!\mathrm{d} S$ is theoretically identical to  $M_0 = \mu A  D$. 
We have verified that their results are the same as the amplitude at the low-frequency plateau of the moment rate spectrum 
\begin{equation}\label{eqn:OmegaFourier}
    \Omega(f) = \mathcal{F}(\dot{M}(t)).
\end{equation}
where $\mathcal{F}$ denotes Fourier transformation.

\subsection*{Energy Calculations}
The total strain energy released from an arrested earthquake rupture can be integrated solely on the ruptured domain $\Sigma$\cite{Kostr1976}:
\begin{equation}
    \Delta W = \frac{1}{2}\int_\Sigma\left[ \tau_\mathrm{{i}}(x, z) +  \tau_\mathrm{{f}}(x, z)\right]\delta_\mathrm{f}(x, z) \mathop{}\!\mathrm{d} S,
\end{equation}
which can be rewritten into a simpler form with the energy-considered averaging method\cite{Noda2012,Noda2013}:
\begin{equation}\label{eqn:DWA}
    \Delta W / A= \frac{1}{2} \left(\overline{{{\tau}}}_\mathrm{{i}}^\mathrm{{E}} + \overline{{{\tau}}}_\mathrm{{f}}^\mathrm{{E}}\right) D~,
\end{equation}
where $\overline{{{\tau}}}_\mathrm{{}}^\mathrm{{E}}$ is the average stress weighted by the final slip
\begin{equation}\label{eqn:energy-considered-averaged-stress}
    \overline{{{\tau}}}_\mathrm{{}}^\mathrm{{E}}(t) = \frac{\int_\Sigma \tau(x, z, t)\delta_\mathrm{f}(x, z) \mathop{}\!\mathrm{d} S}{\int_\Sigma \delta_\mathrm{f}(x, z) \mathop{}\!\mathrm{d} S}.
\end{equation}
Therefore, the energy-considered averaged stress drop is simply
\begin{equation}\label{eqn:DeltatauE}
    \overline{\Delta\tau}^\mathrm{{E}}= \overline{{{\tau}}}_\mathrm{{i}}^\mathrm{{E}} - \overline{{{\tau}}}_\mathrm{{f}}^\mathrm{{E}},
\end{equation}
where $\overline{{{\tau}}}_\mathrm{{i}}^\mathrm{{E}} = \overline{{{\tau}}}_\mathrm{{}}^\mathrm{{E}}(t=0)$ and $\overline{{{\tau}}}_\mathrm{{f}}^\mathrm{{E}} = \overline{{{\tau}}}_\mathrm{{}}^\mathrm{{E}}(t= t_\mathrm{end})$.

We consider two approaches for radiated energy. First, we compute $ E_\mathrm{{R}}^\mathrm{{N}}$ from integrated on-fault stress and slip evolution.
As shown by ~ref.\cite{Kostr1976} and Appendix C in ~ref.\cite{Ripperger2007}, 
$ E_\mathrm{{R}}$ can be conveniently evaluated numerically from the result of numerical simulation through
\begin{equation}\label{eqn:ERN}
     E_\mathrm{{R}}^\mathrm{{N}} = \frac{1}{2}\int_U \left[ \tau_\mathrm{{f}}(x, z) -  \tau_\mathrm{{i}}(x, z)\right]\delta_\mathrm{f}(x, z) \mathop{}\!\mathrm{d} S - \int_0^{ t_\mathrm{end}}\int_U \left[\tau(x, z, t) -  \tau_\mathrm{{i}}(x, z)\right]\dot{\delta}(x, z, t) \mathop{}\!\mathrm{d} S \mathop{}\!\mathrm{d} t~. 
\end{equation}
This approach is equivalent to an energy conservation equation (following ~ref.\cite{Abercrombie2005}) 
\begin{equation}\label{eqn:ERC}
     E_\mathrm{{R}}^\mathrm{{C}} = \Delta W -  E_\mathrm{{D}}, 
\end{equation}
if the $ \tau_\mathrm{{i}}$ term in the second integration of Eqn.~\ref{eqn:ERN} is reorganized into
\begin{equation}
    \int_0^{ t_\mathrm{end}}\int_U \tau_\mathrm{{i}}(x, z)\dot{\delta}(x, z, t) \mathop{}\!\mathrm{d} S  \mathop{}\!\mathrm{d} t = \int_U \tau_\mathrm{{i}}(x, z)\delta_\mathrm{f}(x, z) \mathop{}\!\mathrm{d} S, 
\end{equation}
and moved into the first term. %
Both approaches yield similar results with some differences for small $\chi$, which we associate to small ruptures arresting before reaching Rayleigh wave speed and the effects of rupture initiation process in our numerical simulation. We use $ E_\mathrm{{R}}^\mathrm{{N}}$ to represent $ E_\mathrm{{R}}$ due to its superior numerical stability (see Supplementary Text S3).

Here we start from the energy conservation equation divided by $A$ on both sides, 
\begin{equation}\label{eqn:EB3}
    \Delta W / A = \overline{G}^\prime +  E_\mathrm{{H}} / A +  E_\mathrm{{R}}/A~,
\end{equation}
where $\Delta W$ is the total strain energy released, $A$ is rupture area, $\overline{G}^\prime$ is the spatially averaged breakdown energy, $ E_\mathrm{{H}}$ is heat, and $ E_\mathrm{{R}}$ is radiated energy. 
Replacing $\Delta W /A = \frac{1}{2}\left(\overline{{{\tau}}}_\mathrm{{i}}^\mathrm{{E}} + \overline{{{\tau}}}_\mathrm{{f}}^\mathrm{{E}}\right) D$\cite{Noda2012,Noda2013}, $ E_\mathrm{{H}}/A=\overline{\Delta\tau}^\mathrm{{E}}_\mathrm{f} D$, and $A = M_0/\left(\mu D\right)$, the equation becomes
\begin{equation}\label{eqn:EB2}
    \frac{1}{2}\left(\overline{{{\tau}}}_\mathrm{{i}}^\mathrm{{E}} + \overline{{{\tau}}}_\mathrm{{f}}^\mathrm{{E}}\right) D = \overline{G}^\prime + \overline{\Delta\tau}^\mathrm{{E}}_\mathrm{f} D +  E_\mathrm{{R}}\frac{\mu D}{M_0}~.
\end{equation}
With some rearrangements and replacing $\overline{{{\tau}}}_\mathrm{{i}}^\mathrm{{E}} - \overline{{{\tau}}}_\mathrm{{f}}^\mathrm{{E}}=\overline{\Delta\tau}^\mathrm{{E}}$, $\overline{G}^\prime$ can be expressed as
\begin{equation}\label{eqn:Gb}
    \overline{G}^\prime = \frac{1}{2}\left(\overline{\Delta\tau}^\mathrm{{E}} - \frac{2\mu E_\mathrm{{R}}}{M_0}\right) D~,
\end{equation}
which takes a very similar form as $ G^\prime$ (Eqn.~\ref{eqn:Gp}).
The deduction above actually assumes that  $\overline{G}^\prime$ is 
\begin{equation}\label{eqn:Gint2}
    \overline{G}^\prime = \int_U\int_0^{\delta_\mathrm{f}}\left(\tau - \tau_\mathrm{{f}}\right)\mathop{}\!\mathrm{d}\delta\mathop{}\!\mathrm{d} S / A~
\end{equation}
by the definition of $ E_\mathrm{{H}}/A=\overline{{{\tau}}}_\mathrm{{f}}^\mathrm{{E}} D =  E_\mathrm{{D}}/A - \overline{G}^\prime$, where $ E_\mathrm{{D}}=\int_U\int_0^{\delta_\mathrm{f}}\tau\mathop{}\!\mathrm{d}\delta\mathop{}\!\mathrm{d} S$ is the dissipated energy. Whereas the fracture energy should be the area above $ \tau_\mathrm{{r}}$,
\begin{equation}\label{eqn:Gint}
    G = \int_U\int_0^{\delta_\mathrm{f}}\left(\tau - \tau_\mathrm{{r}}\right)\mathop{}\!\mathrm{d}\delta\mathop{}\!\mathrm{d} S / A~.
\end{equation}
Similar to ~ref.\cite{Abercrombie2005}, when assuming $G$ and $ \tau_\mathrm{{r}}$ are constants, $\overline{G}^\prime$ can be expressed as 
\begin{equation}\label{eqn:GB}
    \overline{G}^\prime = G + \left( \tau_\mathrm{{r}} - \overline{{{\tau}}}_\mathrm{{f}}^\mathrm{{E}}\right) D~.
\end{equation}
Note that the difference between $\overline{G}^\prime$ and $G$ involves the final-slip-weighted-average final stress $\overline{{{\tau}}}_\mathrm{{f}}^\mathrm{{E}}$, \textit{i.e.}, 
\begin{equation}\label{eqn:energy-considered-averaging}
    \overline{{{\tau}}}_\mathrm{{f}}^\mathrm{{E}} = \frac{\int_\Sigma  \tau_\mathrm{{f}}(x, z)\delta_\mathrm{f}(x, z) \mathop{}\!\mathrm{d} S}{\int_\Sigma \delta_\mathrm{f}(x, z) \mathop{}\!\mathrm{d} S}~,
\end{equation}
and the stress overshoot $\left( \tau_\mathrm{{r}} -  \tau_\mathrm{{f}}(x, z)\right)$ in our models appear to be larger at locations with larger slip, as shown in Supplementary Fig.~S7. This highlights the spatially averaged stress overshoot cannot be used when evaluating the accuracy of $ G^\prime$, as the effect of stress overshoot is clearly amplified by its correlation with larger slip.
 
\begin{addendum}
 \item This research was supported by the National Science Foundation under grant EAR-1763499. 
 \item[Authors contributions] D.S.K, and G.C.M. devised the study; C.-Y.K. performed the numerical modeling; C.-Y.K., D.S.K., and G.C.M. analyzed the data; C.-Y.K. prepared the data archive. C.-Y.K., D.S.K., and G.C.M. wrote the manuscript..
 \item[Competing Interests] The authors declare that they have no competing financial interests.
 \item[Correspondence] Correspondence and requests for materials should be addressed to David S. Kammer~(email: dkammer@ethz.ch).
\end{addendum}

\bibliographystyle{naturemag}

\begin{thebibliography}{10}
\expandafter\ifx\csname url\endcsname\relax
  \def\url#1{\texttt{#1}}\fi
\expandafter\ifx\csname urlprefix\endcsname\relax\def\urlprefix{URL }\fi
\providecommand{\bibinfo}[2]{#2}
\providecommand{\eprint}[2][]{\url{#2}}

\bibitem{Abercrombie2005}
\bibinfo{author}{Abercrombie, R.~E.} \& \bibinfo{author}{Rice, J.~R.}
\newblock \bibinfo{title}{{Can observations of earthquake scaling constrain
  slip weakening?}}
\newblock \emph{\bibinfo{journal}{Geophysical Journal International}}
  \textbf{\bibinfo{volume}{162}}, \bibinfo{pages}{406--424}
  (\bibinfo{year}{2005}).

\bibitem{Tinti2005}
\bibinfo{author}{Tinti, E.}, \bibinfo{author}{Spudich, P.} \&
  \bibinfo{author}{Cocco, M.}
\newblock \bibinfo{title}{{Earthquake fracture energy inferred from kinematic
  rupture models on extended faults}}.
\newblock \emph{\bibinfo{journal}{Journal of Geophysical Research: Solid
  Earth}} \textbf{\bibinfo{volume}{110}}, \bibinfo{pages}{1--25}
  (\bibinfo{year}{2005}).

\bibitem{Rice2006}
\bibinfo{author}{Rice, J.~R.}
\newblock \bibinfo{title}{{Heating and weakening of faults during earthquake
  slip}}.
\newblock \emph{\bibinfo{journal}{Journal of Geophysical Research: Solid
  Earth}} \textbf{\bibinfo{volume}{111}}, \bibinfo{pages}{1--29}
  (\bibinfo{year}{2006}).

\bibitem{Viesca2015}
\bibinfo{author}{Viesca, R.~C.} \& \bibinfo{author}{Garagash, D.~I.}
\newblock \bibinfo{title}{{Ubiquitous weakening of faults due to thermal
  pressurization}}.
\newblock \emph{\bibinfo{journal}{Nature Geoscience}}
  \textbf{\bibinfo{volume}{8}}, \bibinfo{pages}{875--879}
  (\bibinfo{year}{2015}).

\bibitem{Perry2020}
\bibinfo{author}{Perry, S.~M.}, \bibinfo{author}{Lambert, V.} \&
  \bibinfo{author}{Lapusta, N.}
\newblock \bibinfo{title}{{Nearly Magnitude-Invariant Stress Drops in Simulated
  Crack-Like Earthquake Sequences on Rate-and-State Faults with Thermal
  Pressurization of Pore Fluids}}.
\newblock \emph{\bibinfo{journal}{Journal of Geophysical Research: Solid
  Earth}} \textbf{\bibinfo{volume}{125}} (\bibinfo{year}{2020}).

\bibitem{Lambert2020}
\bibinfo{author}{Lambert, V.} \& \bibinfo{author}{Lapusta, N.}
\newblock \bibinfo{title}{{Rupture-dependent breakdown energy in fault models
  with thermo-hydro-mechanical processes}}.
\newblock \emph{\bibinfo{journal}{Solid Earth}} \textbf{\bibinfo{volume}{11}},
  \bibinfo{pages}{2283--2302} (\bibinfo{year}{2020}).

\bibitem{Nielsen2016a}
\bibinfo{author}{Nielsen, S.} \emph{et~al.}
\newblock \bibinfo{title}{{Scaling in natural and laboratory earthquakes}}.
\newblock \emph{\bibinfo{journal}{Geophysical Research Letters}}
  \textbf{\bibinfo{volume}{43}}, \bibinfo{pages}{1504--1510}
  (\bibinfo{year}{2016}).

\bibitem{Lapusta2000}
\bibinfo{author}{Lapusta, N.}, \bibinfo{author}{Rice, J.~R.},
  \bibinfo{author}{Ben-Zion, Y.} \& \bibinfo{author}{Zheng, G.}
\newblock \bibinfo{title}{{Elastodynamic analysis for slow tectonic loading
  with spontaneous rupture episodes on faults with rate- and state-dependent
  friction}}.
\newblock \emph{\bibinfo{journal}{Journal of Geophysical Research: Solid
  Earth}} \textbf{\bibinfo{volume}{105}}, \bibinfo{pages}{23765--23789}
  (\bibinfo{year}{2000}).
\newblock \eprint{0402594v3}.

\bibitem{Barbot2019}
\bibinfo{author}{Barbot, S.}
\newblock \bibinfo{title}{{Slow-slip, slow earthquakes, period-two cycles, full
  and partial ruptures, and deterministic chaos in a single asperity fault}}.
\newblock \emph{\bibinfo{journal}{Tectonophysics}}
  \textbf{\bibinfo{volume}{768}}, \bibinfo{pages}{228171}
  (\bibinfo{year}{2019}).

\bibitem{Ke2021}
\bibinfo{author}{Ke, C.-Y.}, \bibinfo{author}{McLaskey, G.~C.} \&
  \bibinfo{author}{Kammer, D.~S.}
\newblock \bibinfo{title}{{The earthquake arrest zone}}.
\newblock \emph{\bibinfo{journal}{Geophysical Journal International}}
  \textbf{\bibinfo{volume}{224}}, \bibinfo{pages}{581--589}
  (\bibinfo{year}{2021}).

\bibitem{Kammer2019}
\bibinfo{author}{Kammer, D.~S.} \& \bibinfo{author}{McLaskey, G.~C.}
\newblock \bibinfo{title}{{Fracture energy estimates from large-scale
  laboratory earthquakes}}.
\newblock \emph{\bibinfo{journal}{Earth and Planetary Science Letters}}
  \textbf{\bibinfo{volume}{511}}, \bibinfo{pages}{36--43}
  (\bibinfo{year}{2019}).

\bibitem{Svetlizky2017a}
\bibinfo{author}{Svetlizky, I.}, \bibinfo{author}{Kammer, D.~S.},
  \bibinfo{author}{Bayart, E.}, \bibinfo{author}{Cohen, G.} \&
  \bibinfo{author}{Fineberg, J.}
\newblock \bibinfo{title}{{Brittle Fracture Theory Predicts the Equation of
  Motion of Frictional Rupture Fronts}}.
\newblock \emph{\bibinfo{journal}{Physical Review Letters}}
  \textbf{\bibinfo{volume}{118}}, \bibinfo{pages}{125501}
  (\bibinfo{year}{2017}).

\bibitem{Chounet2018}
\bibinfo{author}{Chounet, A.}, \bibinfo{author}{Vall{\'{e}}e, M.},
  \bibinfo{author}{Causse, M.} \& \bibinfo{author}{Courboulex, F.}
\newblock \bibinfo{title}{{Global catalog of earthquake rupture velocities
  shows anticorrelation between stress drop and rupture velocity}}.
\newblock \emph{\bibinfo{journal}{Tectonophysics}}
  \textbf{\bibinfo{volume}{733}}, \bibinfo{pages}{148--158}
  (\bibinfo{year}{2018}).

\bibitem{Ke2018}
\bibinfo{author}{Ke, C.-Y.}, \bibinfo{author}{McLaskey, G.~C.} \&
  \bibinfo{author}{Kammer, D.~S.}
\newblock \bibinfo{title}{{Rupture Termination in Laboratory‐Generated
  Earthquakes}}.
\newblock \emph{\bibinfo{journal}{Geophysical Research Letters}}
  \textbf{\bibinfo{volume}{45}}, \bibinfo{pages}{12784--12792}
  (\bibinfo{year}{2018}).

\bibitem{Eshelby1957}
\bibinfo{author}{Eshelby, J.~D.}
\newblock \bibinfo{title}{{The determination of the elastic field of an
  ellipsoidal inclusion, and related problems}}.
\newblock \emph{\bibinfo{journal}{Proceedings of the Royal Society of London.
  Series A. Mathematical and Physical Sciences}}
  \textbf{\bibinfo{volume}{241}}, \bibinfo{pages}{376--396}
  (\bibinfo{year}{1957}).

\bibitem{Noda2012}
\bibinfo{author}{Noda, H.} \& \bibinfo{author}{Lapusta, N.}
\newblock \bibinfo{title}{{On averaging interface response during dynamic
  rupture and energy partitioning diagrams for earthquakes}}.
\newblock \emph{\bibinfo{journal}{Journal of Applied Mechanics, Transactions
  ASME}} \textbf{\bibinfo{volume}{79}}, \bibinfo{pages}{1--12}
  (\bibinfo{year}{2012}).

\bibitem{Noda2013}
\bibinfo{author}{Noda, H.}, \bibinfo{author}{Lapusta, N.} \&
  \bibinfo{author}{Kanamori, H.}
\newblock \bibinfo{title}{{Comparison of average stress drop measures for
  ruptures with heterogeneous stress change and implications for earthquake
  physics}}.
\newblock \emph{\bibinfo{journal}{Geophysical Journal International}}
  \textbf{\bibinfo{volume}{193}}, \bibinfo{pages}{1691--1712}
  (\bibinfo{year}{2013}).

\bibitem{Ide2001}
\bibinfo{author}{Ide, S.} \& \bibinfo{author}{Beroza, G.~C.}
\newblock \bibinfo{title}{{Does apparent stress vary with earthquake size?}}
\newblock \emph{\bibinfo{journal}{Geophysical Research Letters}}
  \textbf{\bibinfo{volume}{28}}, \bibinfo{pages}{3349--3352}
  (\bibinfo{year}{2001}).

\bibitem{Prieto2004}
\bibinfo{author}{Prieto, G.~A.}, \bibinfo{author}{Shearer, P.~M.},
  \bibinfo{author}{Vernon, F.~L.} \& \bibinfo{author}{Kilb, D.}
\newblock \bibinfo{title}{{Earthquake source scaling and self-similarity
  estimation from stacking P and S spectra}}.
\newblock \emph{\bibinfo{journal}{Journal of Geophysical Research: Solid
  Earth}} \textbf{\bibinfo{volume}{109}}, \bibinfo{pages}{1--13}
  (\bibinfo{year}{2004}).

\bibitem{Allmann2009}
\bibinfo{author}{Allmann, B.~P.} \& \bibinfo{author}{Shearer, P.~M.}
\newblock \bibinfo{title}{{Global variations of stress drop for moderate to
  large earthquakes}}.
\newblock \emph{\bibinfo{journal}{Journal of Geophysical Research: Solid
  Earth}} \textbf{\bibinfo{volume}{114}}, \bibinfo{pages}{1--22}
  (\bibinfo{year}{2009}).

\bibitem{Yoshimitsu2014}
\bibinfo{author}{Yoshimitsu, N.}, \bibinfo{author}{Kawakata, H.} \&
  \bibinfo{author}{Takahashi, N.}
\newblock \bibinfo{title}{Magnitude -7 level earthquakes: A new lower limit of
  self-similarity in seismic scaling relationships}.
\newblock \emph{\bibinfo{journal}{Geophysical Research Letters}}
  \textbf{\bibinfo{volume}{41}}, \bibinfo{pages}{4495--4502}
  (\bibinfo{year}{2014}).

\bibitem{Kostr1976}
\bibinfo{author}{Kostrov, V.} \& \bibinfo{author}{Riznichenko, V.}
\newblock \bibinfo{title}{{Seismic moment and energy of earthquakes, and
  seismic flow of rock}}.
\newblock \emph{\bibinfo{journal}{International Journal of Rock Mechanics and
  Mining Sciences {\&} Geomechanics Abstracts}} \textbf{\bibinfo{volume}{13}},
  \bibinfo{pages}{A4} (\bibinfo{year}{1976}).

\bibitem{Convers2011}
\bibinfo{author}{Convers, J.~A.} \& \bibinfo{author}{Newman, A.~V.}
\newblock \bibinfo{title}{{Global evaluation of large earthquake energy from
  1997 through mid-2010}}.
\newblock \emph{\bibinfo{journal}{Journal of Geophysical Research}}
  \textbf{\bibinfo{volume}{116}}, \bibinfo{pages}{B08304}
  (\bibinfo{year}{2011}).

\bibitem{Baltay2014}
\bibinfo{author}{Baltay, A.~S.}, \bibinfo{author}{Beroza, G.~C.} \&
  \bibinfo{author}{Ide, S.}
\newblock \bibinfo{title}{{Radiated Energy of Great Earthquakes from
  Teleseismic Empirical Green's Function Deconvolution}}.
\newblock \emph{\bibinfo{journal}{Pure and Applied Geophysics}}
  \textbf{\bibinfo{volume}{171}}, \bibinfo{pages}{2841--2862}
  (\bibinfo{year}{2014}).

\bibitem{Denolle2016}
\bibinfo{author}{Denolle, M.~A.} \& \bibinfo{author}{Shearer, P.~M.}
\newblock \bibinfo{title}{{New perspectives on self-similarity for shallow
  thrust earthquakes}}.
\newblock \emph{\bibinfo{journal}{Journal of Geophysical Research: Solid
  Earth}} \textbf{\bibinfo{volume}{121}}, \bibinfo{pages}{6533--6565}
  (\bibinfo{year}{2016}).

\bibitem{Walter2006}
\bibinfo{author}{Walter, W.~R.}, \bibinfo{author}{Mayeda, K.},
  \bibinfo{author}{Gok, R.} \& \bibinfo{author}{Hofstetter, A.}
\newblock \emph{\bibinfo{title}{{The Scaling of Seismic Energy With Moment:
  Simple Models Compared With Observations}}}, \bibinfo{pages}{25--41}
  (\bibinfo{publisher}{American Geophysical Union (AGU)},
  \bibinfo{year}{2006}).

\bibitem{Antolik2004}
\bibinfo{author}{Antolik, M.}
\newblock \bibinfo{title}{{The 14 November 2001 Kokoxili (Kunlunshan), Tibet,
  Earthquake: Rupture Transfer through a Large Extensional Step-Over}}.
\newblock \emph{\bibinfo{journal}{Bulletin of the Seismological Society of
  America}} \textbf{\bibinfo{volume}{94}}, \bibinfo{pages}{1173--1194}
  (\bibinfo{year}{2004}).

\bibitem{Beeler2006}
\bibinfo{author}{Beeler, N.~M.}
\newblock \emph{\bibinfo{title}{{Inferring earthquake source properties from
  laboratory observations and the scope of lab contributions to source
  physics}}}, vol. \bibinfo{volume}{170}, \bibinfo{pages}{99--119}
  (\bibinfo{publisher}{American Geophysical Union (AGU)},
  \bibinfo{year}{2006}).

\bibitem{Prieto2013}
\bibinfo{author}{Prieto, G.~A.} \emph{et~al.}
\newblock \bibinfo{title}{{Seismic evidence for thermal runaway during
  intermediate-depth earthquake rupture}}.
\newblock \emph{\bibinfo{journal}{Geophysical Research Letters}}
  \textbf{\bibinfo{volume}{40}}, \bibinfo{pages}{6064--6068}
  (\bibinfo{year}{2013}).

\bibitem{Ye2016}
\bibinfo{author}{Ye, L.}, \bibinfo{author}{Lay, T.}, \bibinfo{author}{Kanamori,
  H.} \& \bibinfo{author}{Rivera, L.}
\newblock \bibinfo{title}{Rupture characteristics of major and great ($m_w \geq
  7.0$) megathrust earthquakes from 1990 to 2015: 1. source parameter scaling
  relationships}.
\newblock \emph{\bibinfo{journal}{Journal of Geophysical Research: Solid
  Earth}} \textbf{\bibinfo{volume}{121}}, \bibinfo{pages}{826--844}
  (\bibinfo{year}{2016}).

\bibitem{Madariaga1976}
\bibinfo{author}{Madariaga, R.}
\newblock \bibinfo{title}{{Dynamics of an expanding circular fault}}.
\newblock \emph{\bibinfo{journal}{Bulletin of the Seismological Society of
  America}} \textbf{\bibinfo{volume}{66}}, \bibinfo{pages}{639--666}
  (\bibinfo{year}{1976}).

\bibitem{Nielsen2016}
\bibinfo{author}{Nielsen, S.} \emph{et~al.}
\newblock \bibinfo{title}{{G: Fracture energy, friction and dissipation in
  earthquakes}}.
\newblock \emph{\bibinfo{journal}{Journal of Seismology}}
  \textbf{\bibinfo{volume}{20}}, \bibinfo{pages}{1187--1205}
  (\bibinfo{year}{2016}).

\bibitem{Guatteri2000}
\bibinfo{author}{Guatteri, M.} \& \bibinfo{author}{Spudich, P.}
\newblock \bibinfo{title}{{What Can Strong-Motion Data Tell Us about
  Slip-Weakening Fault-Friction Laws?}}
\newblock \emph{\bibinfo{journal}{Bulletin of the Seismological Society of
  America}} \textbf{\bibinfo{volume}{90}}, \bibinfo{pages}{98--116}
  (\bibinfo{year}{2000}).

\bibitem{McLaskey2015}
\bibinfo{author}{McLaskey, G.~C.}, \bibinfo{author}{Kilgore, B.~D.} \&
  \bibinfo{author}{Beeler, N.~M.}
\newblock \bibinfo{title}{{Slip-pulse rupture behavior on a 2 m granite
  fault}}.
\newblock \emph{\bibinfo{journal}{Geophysical Research Letters}}
  \textbf{\bibinfo{volume}{42}}, \bibinfo{pages}{7039--7045}
  (\bibinfo{year}{2015}).

\bibitem{Liao2014}
\bibinfo{author}{Liao, Z.}, \bibinfo{author}{Chang, J.~C.} \&
  \bibinfo{author}{Reches, Z.}
\newblock \bibinfo{title}{{Fault strength evolution during high velocity
  friction experiments with slip-pulse and constant-velocity loading}}.
\newblock \emph{\bibinfo{journal}{Earth and Planetary Science Letters}}
  \textbf{\bibinfo{volume}{406}}, \bibinfo{pages}{93--101}
  (\bibinfo{year}{2014}).

\bibitem{Lambert2021}
\bibinfo{author}{Lambert, V.}, \bibinfo{author}{Lapusta, N.} \&
  \bibinfo{author}{Perry, S.~M.}
\newblock \bibinfo{title}{{Propagation of large earthquakes as self-healing
  pulses or mild cracks}}.
\newblock \emph{\bibinfo{journal}{Nature}} \textbf{\bibinfo{volume}{591}},
  \bibinfo{pages}{252--258} (\bibinfo{year}{2021}).

\bibitem{Sato1973}
\bibinfo{author}{Sato, T.} \& \bibinfo{author}{Hirasawa, T.}
\newblock \bibinfo{title}{{Body wave spectra from propagating shear cracks}}.
\newblock \emph{\bibinfo{journal}{Journal of Physics of the Earth}}
  \textbf{\bibinfo{volume}{21}}, \bibinfo{pages}{415--431}
  (\bibinfo{year}{1973}).

\bibitem{Uchide2010}
\bibinfo{author}{Uchide, T.} \& \bibinfo{author}{Ide, S.}
\newblock \bibinfo{title}{{Scaling of earthquake rupture growth in the
  Parkfield area: Self-similar growth and suppression by the finite seismogenic
  layer}}.
\newblock \emph{\bibinfo{journal}{Journal of Geophysical Research}}
  \textbf{\bibinfo{volume}{115}}, \bibinfo{pages}{B11302}
  (\bibinfo{year}{2010}).

\bibitem{Denolle2019}
\bibinfo{author}{Denolle, M.~A.}
\newblock \bibinfo{title}{{Energetic Onset of Earthquakes}}.
\newblock \emph{\bibinfo{journal}{Geophysical Research Letters}}
  \textbf{\bibinfo{volume}{46}}, \bibinfo{pages}{2458--2466}
  (\bibinfo{year}{2019}).

\bibitem{Meier2017}
\bibinfo{author}{Meier, M.-A.}, \bibinfo{author}{Ampuero, J.~P.} \&
  \bibinfo{author}{Heaton, T.~H.}
\newblock \bibinfo{title}{{The hidden simplicity of subduction megathrust
  earthquakes}}.
\newblock \emph{\bibinfo{journal}{Science}} \textbf{\bibinfo{volume}{357}},
  \bibinfo{pages}{1277--1281} (\bibinfo{year}{2017}).

\bibitem{Beeler2003}
\bibinfo{author}{Beeler, N.~M.}, \bibinfo{author}{Wong, T.~F.} \&
  \bibinfo{author}{Hickman, S.~H.}
\newblock \bibinfo{title}{{On the expected relationships among apparent stress,
  static stress drop, effective shear fracture energy, and efficiency}}.
\newblock \emph{\bibinfo{journal}{Bulletin of the Seismological Society of
  America}} \textbf{\bibinfo{volume}{93}}, \bibinfo{pages}{1381--1389}
  (\bibinfo{year}{2003}).

\bibitem{uguca}
\bibinfo{author}{Kammer, D.~S.}, \bibinfo{author}{Albertini, G.} \&
  \bibinfo{author}{Ke, C.-Y.}
\newblock \bibinfo{title}{{UGUCA: spectral-boundary-integral method for
  modeling fracture and friction}}.
\newblock \bibinfo{howpublished}{\url{https://uguca.gitlab.io/uguca/}}
  (\bibinfo{year}{2021}).
\newblock \bibinfo{note}{[Online; accessed 1-March-2021]}.

\bibitem{Geubelle1995}
\bibinfo{author}{Geubelle, P.~H.}
\newblock \bibinfo{title}{{A spectral method for three-dimensional
  elastodynamic fracture problems}}.
\newblock \emph{\bibinfo{journal}{Journal of the Mechanics and Physics of
  Solids}} \textbf{\bibinfo{volume}{43}}, \bibinfo{pages}{1791--1824}
  (\bibinfo{year}{1995}).

\bibitem{Geubelle1997}
\bibinfo{author}{Geubelle, P.~H.} \& \bibinfo{author}{Breitenfeld, M.~S.}
\newblock \bibinfo{title}{{Numerical analysis of dynamic debonding under
  anti-plane shear loading}}.
\newblock \emph{\bibinfo{journal}{International Journal of Fracture}}
  \textbf{\bibinfo{volume}{85}}, \bibinfo{pages}{265--282}
  (\bibinfo{year}{1997}).

\bibitem{Breitenfeld1998}
\bibinfo{author}{Breitenfeld, M.~S.} \& \bibinfo{author}{Geubelle, P.~H.}
\newblock \bibinfo{title}{{Numerical analysis of dynamic debonding under 2D
  in-plane and 3D loading}}.
\newblock \emph{\bibinfo{journal}{International Journal of Fracture}}
  \textbf{\bibinfo{volume}{93}}, \bibinfo{pages}{13--38}
  (\bibinfo{year}{1998}).

\bibitem{Harris2018}
\bibinfo{author}{Harris, R.~A.} \emph{et~al.}
\newblock \bibinfo{title}{{A Suite of Exercises for Verifying Dynamic
  Earthquake Rupture Codes}}.
\newblock \emph{\bibinfo{journal}{Seismological Research Letters}}
  \textbf{\bibinfo{volume}{89}} (\bibinfo{year}{2018}).

\bibitem{Ripperger2007}
\bibinfo{author}{Ripperger, J.}, \bibinfo{author}{Ampuero, J.-P.},
  \bibinfo{author}{Mai, P.~M.} \& \bibinfo{author}{Giardini, D.}
\newblock \bibinfo{title}{{Earthquake source characteristics from dynamic
  rupture with constrained stochastic fault stress}}.
\newblock \emph{\bibinfo{journal}{Journal of Geophysical Research: Solid
  Earth}} \textbf{\bibinfo{volume}{112}}, \bibinfo{pages}{1--17}
  (\bibinfo{year}{2007}).

\end{thebibliography}

\end{document}